# 28 OCTOBER 2003 FLARE: HIGH- ENERGY GAMMA EMISSION, TYPE II RADIO EMISSION AND SOLAR PARTICLE OBSERVATIONS

S. N. Kuznetsov, V. G. Kurt*, B. Yu. Yushkov, I. N. Myagkova,

*Skobeltsyn Institute of Nuclear Physics of Moscow State University, Moscow 119992, Russia*

K. Kudela

*Institute of Experimental Physics, Slovak Academy of Sciences SK-04353 Kosice, Slovak Republic*

A. V. Belov,

*Institute of Terrestrial Magnetism, Ionosphere and Radio Wave Propagation Troitsk, Moscow Region, 142092, Russia*

C. Caroubalos, A. Hilaris, H. Mavromichalaki, X. Moussas **,
and P. Preka-Papadema

*Physics Department, University of Athens, 15571, Athens, Greece*

The 28 October 2003 flare gave us the unique opportunity to compare the acceleration time of high- energy protons with the escaping time of those particles which have been measured onboard spacecraft and by neutron monitors network as GLE event. High-energy emission time scale and shock wave height and velocity time dependencies were also studied.

The experiment SONG (SOlar Neutrons and Gamma-rays) is designed for the detection of X-ray and γ-ray emission in the energy range $0.05 \div 100$ MeV, neutrons with $E_n > 20$ MeV, electrons with $E_e = 12 \div 100$ MeV, as well as protons with $E_p > 75$ MeV. Gamma ray emission with $E_\gamma > 60$ MeV observed by SONG onboard CORONAS-F in the 28 October 2003 major flare X 17.2/4B (S16, E08) provided an evidence for the relativistic particle appearance during the flare development. Consequently we got a unique opportunity to compare the acceleration time to the escaping time of the protons with $E_p > 200$ MeV energies measured as GLE and onboard the Spacecraft.

The flare impulsive phase start time was defined as $T_0 = 60 \div 120$ s after 11:00:00 UT according to the measurements in the various wavelengths. Hereafter zero of a time scale corresponds to 11:00:00 UT. Figure 1 (left bottom panel) shows the most representative response of SONG detector to the γ-ray emission in $7 \div 15$ MeV and $60 \div 100$ MeV energy ranges. The first 60 s of the high-energy emission correspond to the most bright outburst for the energy ≤ 40 MeV. Subsequent delayed emission began suddenly at 225 s and extends more than 400 s with the photon energy reached 100 MeV. The right top panel of Fig. 1 demonstrates that the first impulsive flare phase is characterized by a mainly bremsstrahlung spectrum (generated by electrons) with a small excess in the γ-line range 0.5 - 6 MeV (γ-line produced by protons with $E_p = 10 \div 30$ MeV). The spectrum of the delayed phase has a remarkable plateau in the energy range > 25 MeV. This spectrum irregularity is attributed to the γ-emission resulted by $\pi^0$ decay process which needs the protons with $E_p > 200$ MeV. The absence of smooth transition from the impulsive phase to delayed phase for all energy γ-ray emission makes it tempting to assume that two different acceleration mechanisms were in operation causing both impulsive and delayed phases of the flare (Refs. 1, 2).

The question "does a shock wave propagating in the Corona accelerate particles up to ultra-relativistic energies during the delay phase of the flare?" is being discussed briefly. The shock wave formation, its height-time and height-velocity dependencies were calculated based on the data of radiospectrograph ARTEMIS Type II radio emission in the 650 ± 20 MHz range and were compared with the analogous dependencies of CME.3 Com- paring the MHD wave velocity, calculated from the type II frequency drift, with the CME velocity we note that the former is a model dependent calculation, assuming a Newkirk corona and radial propagation; the latter is subject to projection effects. However, the calculated velocities are close enough (see right bottom panel of Fig. 1) to allow us to characterize the type II burst as a shock driven by the front of CME (Ref. 4). The extrapolation of shock wave and CME trajectories back to the Sun surface permitted to define their liftoff time equal to 100 ± 30 s that corresponds to the time of the main flare energy release T0 . The initial velocity of the blob is of $9 \times 10^2 - 1.4 \times 10^3$ km · s$^{-1}$ and its acceleration value is of 0.12 ± 0.3 km · s$^{-2}$ .

The appearance at 188 s of type II burst in the middle corona at 0.2 RSun indicates the time of a supersonic shock wave formation. The shock wave formation precedes the sharp onset of γ-ray emission delayed phase. Perhaps this 30 - 35 s time difference testifies against proton acceleration by the shock wave front moving upward. An alternative possibility of protons acceleration has been proposed in models invoking magnetic field reconnection. The rising CME/shock front and neutral point are connected by a stretched current sheet. According to Ref. 5, protons can be actually accelerated to GeV energies in high temperature turbulent reconnection current sheet during the late phase of solar flare.

Now let us compare the arrival time of proton at the Earth with the time of the γ-ray emission. The McMurdo, Norilsk, Potchefstrom neutron monitors (NM) indicated the earliest onset among NM network at 750 ± 30 s. The arrival time of the protons with energy $E_p$ = 200 ÷ 300 MeV at CORONAS-F is equal to 750 s and coincides with monitors onset time (see Fig. 1, upper left panel). Taking into account that propagation time of γ-ray emission is equal to 508 s we can compare time of γ-ray emission with the escaping time of the protons which arrived to the Earth at 750 s. The shortest possible propagation time values for the protons having zero pitch angle can be estimated as 1000 s for the 200 MeV energy proton (v = 0.566c) and as 890s for the 300 MeV protons (v = 0.652c)in the assumption that their path along the interplanetary magnetic field lines is 1.17 AU. Time delay 500 s and 388 s in respect to γ-ray detection gives us the latest time interval when protons being measured at 1 AU could escape from the Sun. This time interval marked as a shaded box in Fig. 1 (bottom left panel) indicates the 'leakage' time of 200 ÷ 300 MeV protons into the interplanetary space. From the other hand the pion decay γ-ray emission indicates the time of the acceleration of a proton with energy ≥ 300 MeV. Foregoing presents a conclusive proof that the protons started to escape from the Sun in the same moment when the pion decay γ-ray emission was generated without any time delay.

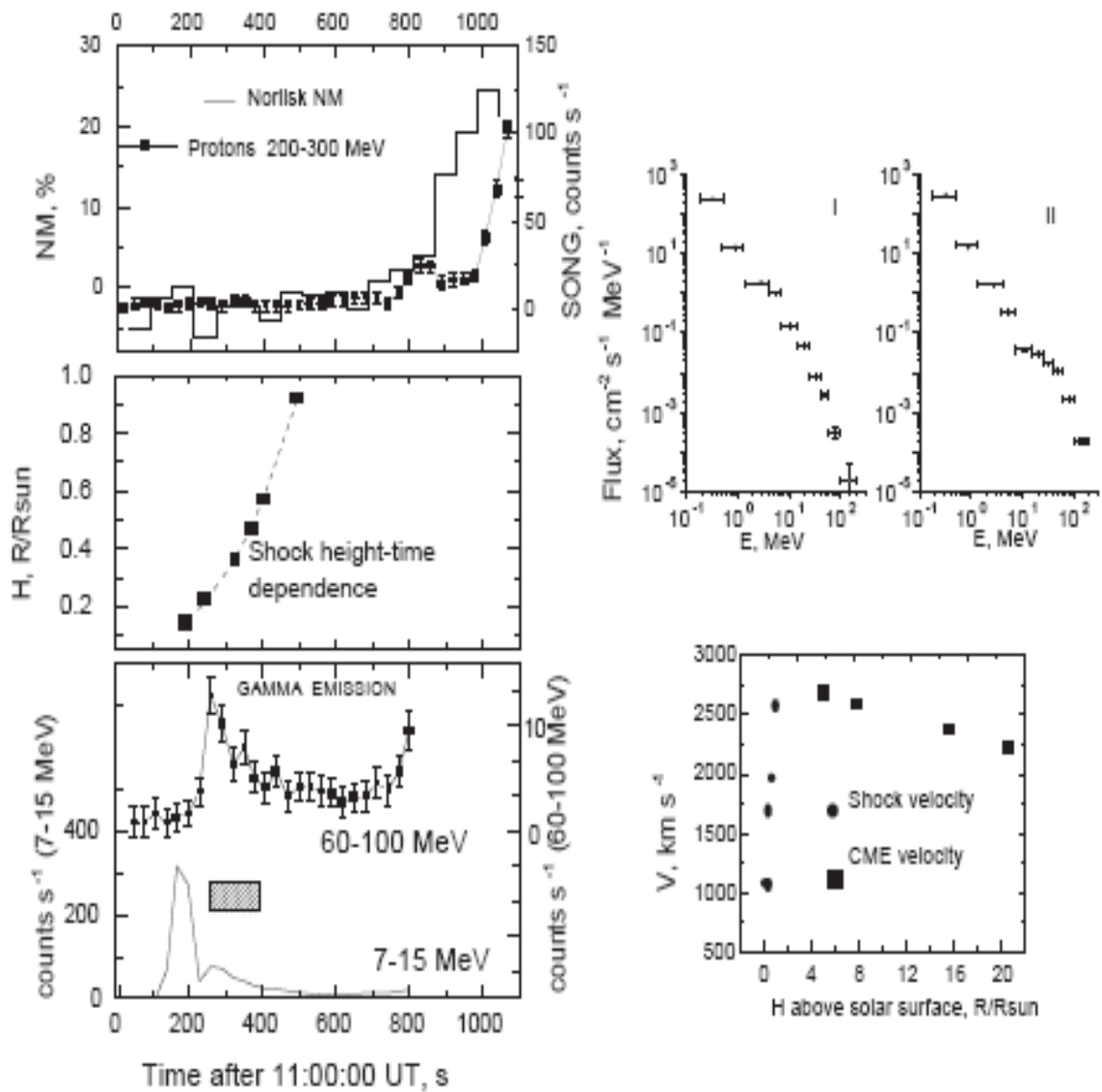

Fig. 1. SONG detector response to the -ray emission in the 7 χ 15 MeV and 60 χ 100 MeV energy ranges (left bottom); protons onset in the 200 χ 300 MeV energy range (left top) and shock wave height (left middle) as functions of time [s]. The -ray emission spectrum shape integrated over impulsive and delayed phases (right top) and Combined shock wave – CME height-velocity dependence (right bottom).